\documentclass{article}
\usepackage{spconf,amsmath,graphicx}

\usepackage[noadjust]{cite}

\usepackage{amsmath,amssymb,bm}
\usepackage{url}
\usepackage{lipsum}

\usepackage{hyperref}
\usepackage[dvipsnames]{xcolor}
\newcommand\myshade{70}
\colorlet{mywholecolor}{MidnightBlue}
\hypersetup{
  linkcolor  = mywholecolor!\myshade!black,
  citecolor  = mywholecolor!\myshade!black,
  urlcolor   = mywholecolor!\myshade!black,
  colorlinks = true,
}

\usepackage{array}
\newcommand{\PreserveBackslash}[1]{\let\temp=\\#1\let\\=\temp}
\newcolumntype{C}[1]{>{\PreserveBackslash\centering}p{#1}}
\newcolumntype{R}[1]{>{\PreserveBackslash\raggedleft}p{#1}}
\newcolumntype{L}[1]{>{\PreserveBackslash\raggedright}p{#1}}

\newcommand{\datasetname}{ASID}
\newcommand{\tabspace}{\vspace{-2mm}}

\usepackage[dvipsnames]{xcolor}

\newcommand\minisec[1]{\vspace{0.8mm}\noindent\textbf{#1 ---}}

\title{Investigating the efficacy of music version retrieval systems for setlist identification}
\name{Furkan Yesiler\thanks{This work is supported by the MIP-Frontiers project, the European Union's Horizon 2020 research and innovation programme under the Marie Skłodowska-Curie grant agreement No.~765068.}$^{\star}$ \qquad Emilio Molina$^{\dagger}$ \qquad Joan Serr{\`a}$^{\ddagger}$ \qquad Emilia G{\'o}mez$^{\dagger\dagger \star}$}
\address{$^{\star}$ Music Technology Group, Universitat Pompeu Fabra, Barcelona, Spain \\
$^{\dagger}$ BMAT Licensing S.L., Barcelona, Spain \qquad
$^{\ddagger}$ Dolby Laboratories, Barcelona, Spain \\
$^{\dagger\dagger}$ Joint Research Centre, European Commission, Sevilla, Spain} 

\begin{document}
\ninept
\maketitle
\begin{abstract}

The setlist identification (SLI) task addresses a music recognition use case where the goal is to retrieve the metadata and timestamps for all the tracks played in live music events. Due to various musical and non-musical changes in live performances, developing automatic SLI systems is still a challenging task that, despite its industrial relevance, has been under-explored in the academic literature. In this paper, we propose an end-to-end workflow that identifies relevant metadata and timestamps of live music performances using a version identification system. We compare 3 of such systems to investigate their suitability for this particular task. For developing and evaluating SLI systems, we also contribute a new dataset that contains 99.5\,h of concerts with annotated metadata and timestamps, along with the corresponding reference set. The dataset is categorized by audio qualities and genres to analyze the performance of SLI systems in different use cases. Our approach can identify 68\% of the annotated segments, with values ranging from 35\% to 77\% based on the genre. Finally, we evaluate our approach against a database of 56.8\,k songs to illustrate the effect of expanding the reference set, where we can still identify 56\% of the annotated segments.

\end{abstract}
\section{Introduction}
\label{sec:intro}


Music recognition is commonly used to refer to the task of identifying the presence of a known music track in an unknown audio stream, ideally along with its start and end timestamps~\cite{Wang2006a,Arcas2017}. While the most common and successful technologies for music recognition are audio fingerprinting systems~\cite{Cano2005} that can identify recordings with slight degradations (e.g.,~background noise~\cite{Haitsma2001}, voiceovers~\cite{Wang2003}, or pitch shifting~\cite{Fenet2011a, Joren2014, Sonnleitner2014}), they tend to perform poorly for live performance monitoring. Live performances can incorporate many alterations from the studio recordings, including changes in tempo, key, structure, background noise, additional applause and banters, and so on. Therefore, identifying live music content typically requires version identification (VI) systems, which are designed to go beyond near-exact duplicate detection to identify recordings that, although having perceptual differences, convey the same musical entity (e.g.,~live performances or cover songs)~\cite{serra2009, bertin2012, doras2019, yesiler2020a, yesiler2020b}.

The setlist identification (SLI) of live music performances (i.e.,~full concerts) stands as a challenging branch within music recognition. In the music information retrieval community, SLI was formally defined by Wang et al.~\cite{wang2014} and divided into two sequential sub-tasks, where the first task aims to retrieve only the related metadata in the correct order, and the second task concerns further processing of the retrieved items to obtain correct timestamps. The main applications of SLI systems include automatic generation of metadata and timestamps for concerts in streaming platforms (e.g.,~Youtube) and copyright management for the music industry. The vast variety of music usage contexts in digital platforms makes it impossible to track music usage manually and, therefore, highlights the necessity of automatic systems.

The work by Wang et al.~\cite{wang2014}, together with a few submissions to MIREX 2015\footnote{\url{https://www.music-ir.org/mirex/wiki/2015:Set_List_Identification}} is, to the best of our knowledge, the only one specifically targeting the SLI task. 
Although some works in the VI literature address the use case of live performance identification~\cite{rafii2014, tsai2017}, the proposed approaches and evaluation contexts do not consider entire concerts nor retrieving timestamps. Therefore, one cannot consider them as examples for SLI. Wang et al.~\cite{wang2014} assume that the artist is known for each concert. For overlapping windows, a VI system is used to return a set of candidates using thumbnails, and the final matches and their boundaries are identified among those candidates using a dynamic time warping algorithm. The system is evaluated on a dataset of 20~concerts from 10~rock bands, using three metrics: edit distance, boundary deviation, and frame accuracy. Although demonstrating a plausible performance, scaling such a system to reference databases of thousands of songs stands as a difficult challenge due to the computational complexities of the algorithms used in each step.

In this paper, we study the efficacy of current VI systems for SLI, considering a range of use cases related to the monitoring of live performances. To mimic a realistic industrial scenario, our approach combines the sub-tasks proposed by Wang et al.~\cite{wang2014} into a single task by creating an end-to-end workflow that takes audio signals as input and creates a final document with the retrieved metadata and timestamps. By using predetermined window and hop sizes, we compare the performances of three VI systems that produce overlapping matches that we further process to create the final results. We develop and evaluate our system using a new dataset of 75~concerts that are categorized by varying audio qualities and genres, and annotated in terms of the songs played in each concert and their timestamps. We study the impact of audio quality, genre, and reference set size (up to 56.8\,k songs) on system performance. We report our findings using 4~evaluation metrics following industrial practices. We publish our dataset and evaluation code at \url{https://github.com/furkanyesiler/setlist_id}.

\begin{figure*}[t]
  \centering
  \includegraphics[width=\textwidth]{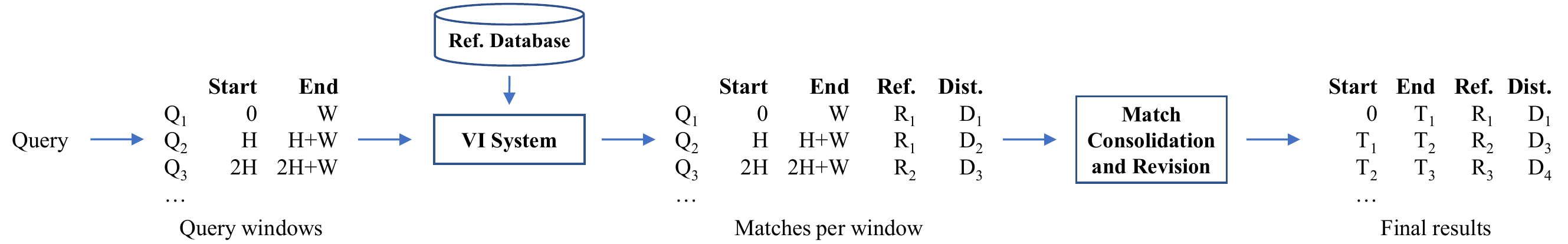}
  \caption{Overall block diagram of our end-to-end workflow.}
  \label{fig:block_diagram}
\end{figure*}

\section{Methodology}
\label{sec:method}

\subsection{System overview}

Our workflow consists of several steps for processing a query (i.e.,~audio file of a concert) and a reference database to retrieve a list of songs and their respective start and end timestamps (Fig.~\ref{fig:block_diagram}). Firstly, we process the audio queries with a sliding window of size $W$ and a hop size $H$ (windowing is not applied to the reference songs). For each windowed query $Q_i$, we use a VI system to retrieve the most similar item from the reference database. After obtaining individual matches for each window, we perform a number of post-processing steps to consolidate and revise those, and form a final list of results. Lastly, we compute several evaluation metrics.

\subsection{Input representation}
The first step of our system is to extract useful information from music audio signals. For this, we use cremaPCP representations~\cite{mcfee2017structured, yesiler2020a}, extracted with the pre-trained model shared in \url{https://github.com/bmcfee/crema}. Recent works in VI show that this pitch class profile variant generally improves system performance, compared to other variants~\cite{yesiler2019}. We use a hop size of 4,096~samples for audio signals sampled at 44.1\,kHz. 

\subsection{VI systems}\label{sec:vi-sys}
For obtaining pairwise distance values between query windows and references, we compare three VI systems. We consider $W= \{120, 180, 240\}$\,s and $H = \{15, 30, 60\}$\,s. 
We perform initial experiments on our development set to pick the best $W$ and $H$ values for each algorithm based on the total length of correctly-identified segments (see DLP metric in Sec.~\ref{sec:eval-met}). Considered systems are:

\minisec{Re-MOVE} Re-MOVE~\cite{yesiler2020b} is a recent model that is trained with embedding distillation techniques to further improve both the accuracy and the scalability aspects of a state-of-the-art VI system~\cite{yesiler2020a}. It encodes each song into an embedding vector of size 256. We transfer the pre-trained weights of the model shared in \url{https://github.com/furkanyesiler/re-move} into an equivalent Keras~\cite{chollet2015keras} model (no re-training or fine-tuning is performed). As distance between embeddings, we use cosine distance.

\minisec{Qmax} Qmax refers to the VI system proposed by Serrà et al.~\cite{serra2009}. The similarity estimation between two songs is performed using a local alignment algorithm, and the length of the longest aligned subsequence is considered as the distance between the songs after being normalized by the length of the reference song. We use the implementation shared in Essentia~\cite{essentia} with the default parameters.

\minisec{2DFTM} 2DFTM refers to the embedding-based VI system proposed by Bertin-Mahieux and Ellis~\cite{bertin2012}. It is one of the first algorithms proposed in the VI literature to address large-scale retrieval scenarios. To use beat-synchronous features as the proposed approach, we perform beat tracking on each query window and reference using onset strength envelopes pre-computed with librosa~\cite{librosa}. As distance between embeddings, we use Euclidean distance.

\subsection{Consolidation and revision of potential matches}

Using the VI systems described above, we compute the distances between each query window $Q_i$ and each item from the reference database. The reference song with the lowest distance to $Q_i$ is considered as its potential match. However, using a windowing schema can create several potential matches for query segments. To reduce the number of matches to a single match for any given time frame, we perform a series of operations to consolidate and revise the obtained potential matches. Note that, although the VI system can be considered as the main component of our entire workflow, this last step is highly important to obtain a useful final list of results. 


Our first step is to merge the consecutive overlapping matches that return the same reference song, and the distance value for the merged match is selected as the lowest distance among the respective matches. Next, to avoid the overlapping matches that return different reference songs, we obtain all possible overlaps and select the reference song that comes from the match with the lowest distance for each overlapping segment. Finally, we perform another merging step by joining any consecutive matches that return the same reference and have no gap between them (i.e.,~the matches that may be split in the previous step). The final results do not contain overlapping matches for any segment of the query. 

Our initial experiments showed that this consolidation and revision step is useful for filtering out many incorrect matches, however, along with a few correct ones. To further reduce the number of incorrect matches, we simply train a support vector machine model for a binary classification (correct/incorrect) task, using the scikit-learn library~\cite{scikit-learn} and the distance and duration values of correct and incorrect matches as features. We see that this simple classifier drastically reduces the number of false positives, however, at the expense of slightly increased false negatives.

\subsection{Dataset}

\begin{table}[tb!]
\begin{center}
\label{tab:ablation}
\begin{tabular}{l c c c | c}
\hline\hline
Genre & AQ-A &  AQ-B & AQ-C & Total \\
\hline

Pop/Commercial & 8 (5)  &  3 & 3 & 14 (5) \\ 
Rock/Metal & 8 (3)  &  7 & 6  & 21 (3) \\ 
Indie/Alternative & 5  &  7 & 3  & 15    \\ 
Hip-hop/Rap & 5 (2)  &  0 & 3  & 8  (2) \\ 
Electronic & 6   &  1 & 0  & 7   \\ 
\hline
Total & 32 (10) &  18 & 15 & 65 (10) \\

\hline\hline
\end{tabular}
\caption{Number of concerts per audio quality and genre. The numbers in parenthesis indicate the concerts in the development set.}\label{tab:data}
\end{center}
\vspace{-3mm}
\end{table}

For our experiments, we have collected and annotated a new dataset, \datasetname: automatic setlist identification dataset. It contains pre-extracted features, metadata, Youtube or Soundcloud links, and timestamp annotations for 75~concerts and all the relevant reference songs (i.e.,~the songs that are played in each concert). Concert durations range between 21.7\,min and 2.5\,h, with a total duration of 99.5\,h. The total number of reference songs is 1,298, with a total duration of 90.1\,h. As mentioned in Sec.~\ref{sec:intro}, we make this dataset publicly available.

\datasetname\ includes a variety of use cases regarding audio quality and genres. For this, we have selected three categories for audio quality: AQ-A, AQ-B, and AQ-C. AQ-A contains high-quality recordings, mainly coming from broadcast recordings or official releases. AQ-B contains professionally recorded concerts, mainly from small venues (in general, we observe that the mixing/mastering quality for concerts in AQ-B is inferior to the ones in AQ-A). Lastly, AQ-C contains smartphone or video camera recordings from varying-size venues/events. In terms of genre, we categorize the concerts into 5~main groups: pop/commercial, rock/metal, indie/alternative, hip-hop/rap, and electronic. The number of concerts for each audio quality and genre can be seen in Table~\ref{tab:data}.

We use 10~concerts (14.3\,h) as a separate development set to select $W$ and $H$ for each VI system, and to train a classifier for the match revision step. The references for the development set include 180~songs. The remaining 65~concerts (85.2\,h) and the related reference set is used for the main results. The total number of annotated segments for the evaluation set is 1,138, with a duration of 80.6\,h.

\subsection{Evaluation metrics}\label{sec:eval-met}

Following common practice in industrial contexts, we evaluate our approach using: (1) true positives (TP), the number of matches (after merging and removing overlaps) that overlap the correct annotations in the ground truth (several correct matches that overlap the same annotation are counted separately); (2) false positives (FP), the number of matches (after merging and removing overlaps) that do not correspond to the related annotation; (3) detected annotations percentage (DAP), the ratio of the number of detected annotations with respect to the total number of annotations in the ground truth; and (4) detected length percentage (DLP), the ratio of the correctly-identified duration of all TP with respect to the total duration of annotations in the ground truth.



\section{Results}
\label{sec:results}

\subsection{Overall results}

Based on the DLP values obtained for the development set, we select $(W,H)$ pairs for each considered VI system. For Re-MOVE and Qmax, we select both (120,30) and (120,60) since the DLP values using those pairs result in very minor differences, and for 2DFTM, we select only (120,15). Due to the computation time needed to evaluate Qmax on our test set (see Sec.~\ref{sec:run-time}), we compute results for only (120,30), and simulate the results for (120,60) by skipping the matches for every second window.

The overall results for each considered system and $(W,H)$ pair show that, while the performances of Re-MOVE and Qmax systems are fairly close, they both outperform 2DFTM by a considerable margin (Table~\ref{tab:overall-res}). Both Re-MOVE and Qmax could identify +78\% of the annotated segments (DAP metric, before the classifier), which is a considerably good performance albeit using arbitrary windows for retrieval instead of clearly-segmented ones. The differences between DAP and DLP values suggest imprecise timestamp retrievals that result mainly from using fixed $W$ and $H$ values without any fine-grained refinements on the timestamp resolution. 

\begin{table}[t]
\begin{center}

\begin{tabular}{l C{1.cm} C{1.1cm} C{1.3cm} C{1.3cm}}
\hline\hline
VI Config. & TP & FP & DAP (\%) & DLP (\%) \\
\hline
R - (120,30) &   \textbf{936}\,/\,\textbf{771} & 1075\,/\,177   &  \textbf{80.3}\,/\,\textbf{67.8} &    60.5\,/\,56.9 \\
R - (120,60) &   922\,/\,735 & \textbf{1013}\,/\,112   &  79.8\,/\,64.6 &    60.3\,/\,54.8 \\
Q - (120,30) &   902\,/\,766 & 1217\,/\,132   &  78.3\,/\,\textbf{67.8} &    60.8\,/\,\textbf{57.5} \\
Q - (120,60) &   905\,/\,736 & 1039\,/\,\textbf{75}    &  78.4\,/\,64.9 &    \textbf{60.9}\,/\,56.0 \\
F - (120,15) &   736\,/\,524 & 2686\,/\,453   &  61.3\,/\,46.0 &    41.1\,/\,37.1 \\

\hline\hline
\end{tabular}
\caption{Overall results for 5 configurations on the evaluation set. R, Q, and F denote Re-MOVE, Qmax, and 2DFTM, respectively. The left/right values denote the metrics before/after the classifier.}\label{tab:overall-res}
\end{center}
\end{table}

\begin{figure}[t]
  \centering
  \includegraphics[width=0.95\linewidth]{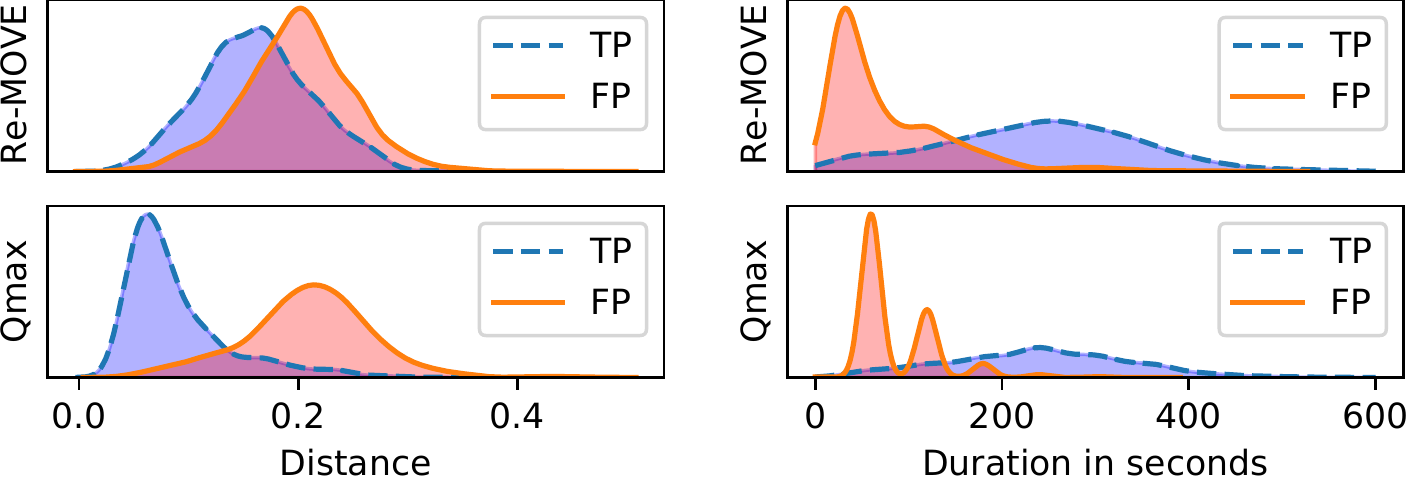}
  \caption{Distance (left) and duration (right) distributions of TP and FP for Re-MOVE (120,30) (top) and Qmax (120,60) (bottom).}
  \label{fig:dist-tpfp}
\end{figure}

\begin{table}[t]
\begin{center}
\begin{tabular}{l C{1.cm} C{1.1cm} C{1.3cm} C{1.3cm}}
\hline\hline
VI Config. & TP & FP & DAP (\%) & DLP (\%) \\
\hline
\multicolumn{5}{c}{\textit{AQ-A}} \\
\hline
R - (120, 30) &   \textbf{480}\,/\,\textbf{412} & 586\,/\,96   &  \textbf{80.6}\,/\,\textbf{71.1} &    55.6\,/\,\textbf{53.2} \\
Q - (120, 60) &   458\,/\,393 & \textbf{572}\,/\,\textbf{47}    &  77.5\,/\,67.6 &    \textbf{55.5}\,/\,52.3 \\

\hline
\multicolumn{5}{c}{\textit{AQ-B}} \\
\hline
R - (120, 30) &   215\,/\,176 & 245\,/\,37   &  87.6\,/\,73.4 &    69.0\,/\,64.2 \\
Q - (120, 60) &   \textbf{221}\,/\,\textbf{179} & \textbf{195}\,/\,\textbf{9}    &  \textbf{89.6}\,/\,\textbf{75.1} &    \textbf{72.6}\,/\,\textbf{66.6} \\

\hline
\multicolumn{5}{c}{\textit{AQ-C}} \\
\hline
R - (120, 30) &   \textbf{241}\,/\,\textbf{183} & \textbf{244}\,/\,44   &  \textbf{74.4}\,/\,\textbf{57.6} &    \textbf{67.3}\,/\,\textbf{61.1} \\
Q - (120, 60) &   226\,/\,164 & 272\,/\,\textbf{19}    &  71.5\,/\,51.9 &    65.8\,/\,56.8 \\

\hline\hline
\end{tabular}
\caption{Results based on audio quality. R and Q denote Re-MOVE and Qmax, respectively. The left/right values denote the metrics before/after the classifier.}\label{tab:aq-results}
\end{center}
\end{table}

The values before and after the classifier show that even a simple classifier can reduce the number of FPs by more than 80\% while reducing the TPs by only 15--20\% (excluding 2DFTM) and the DLPs by 4.5\% on average. This suggests that the distance and duration distributions for TPs and FPs are different enough to enable useful classification, especially for Qmax (120, 60)  (Fig.~\ref{fig:dist-tpfp}).

\subsection{Breakdown of results}

\minisec{Audio quality} 
We now present results separately for each audio quality, using a limited set of configurations (Table~\ref{tab:aq-results}). The total number of annotated segments and the total duration of those (TA and TL, respectively) for categories AQ-A, AQ-B, and AQ-C are (581,~48.8\,h), (241,~15.8\,h), and (316,~16.9\,h), respectively. The results suggest that, surprisingly, the audio quality is not the most crucial factor affecting system performance as both systems result in higher DLP values for AQ-B and AQ-C compared to AQ-A. This shows that our input representation performs robustly against noise. The low values for AQ-A may mainly result from the variety of included genres/styles. 

\begin{figure}[tb!]
  \centering
  \includegraphics[width=0.95\linewidth]{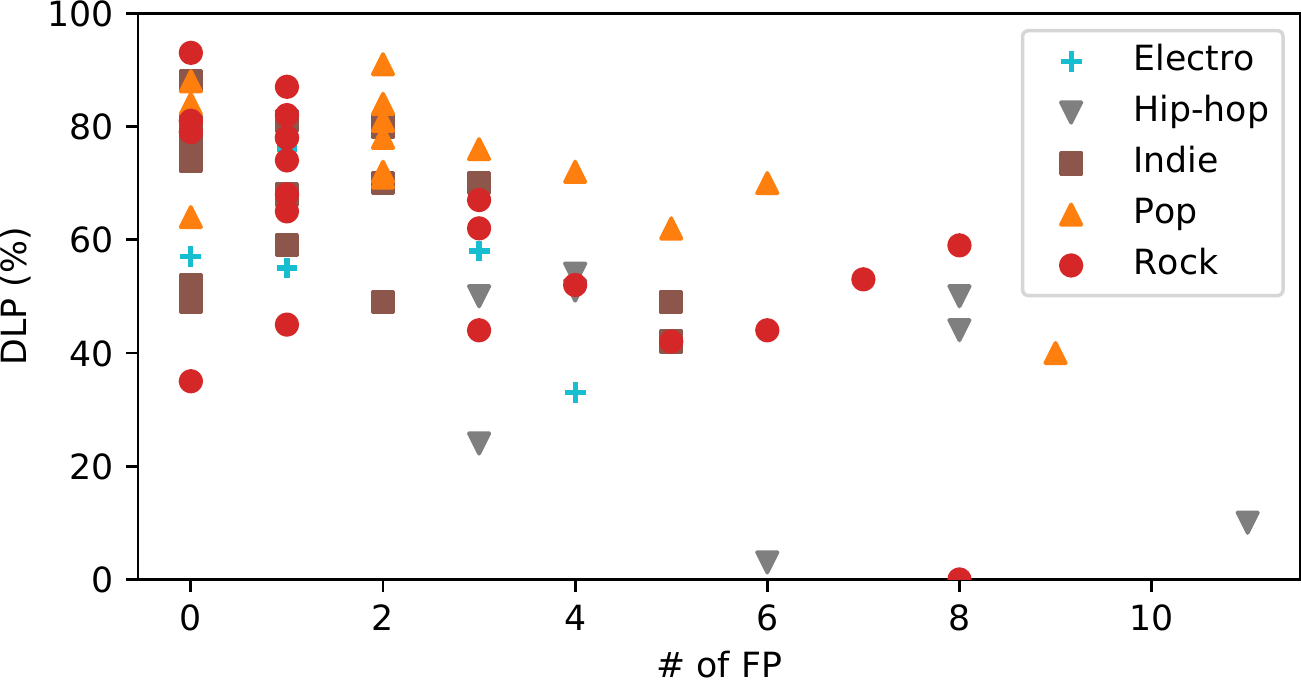}
  \caption{DLP and FP values after the classifer for each concert evaluated with Re-MOVE (120,30), categorized by genre.}
  \label{fig:dlp-fp}
\end{figure}

\minisec{Genre} 
The results categorized by genres can be seen in~Table~\ref{tab:gen-results}. TA and TL values for each category from top to bottom are (272,~17.1\,h), (372,~25.6\,h), (208,~12.3\,h), (171,~17.6\,h) and (115,~8.0\,h), respectively.
We observe that the performances of both VI systems are consistent across genres, with ``Hip-hop/Rap'' being an outlier. Relying on only harmonic information for retrieval leads to a drastic performance decrease for certain musical styles (i.e.,~hip-hop). However, the effect of genre on system performance is not the only decisive factor. The results depicted in Fig.~\ref{fig:dlp-fp} show that the system performance can show a large variance among concerts even within the same genre, with ``Pop/Commercial'' having the most consistent results.

\minisec{Reference set size} 
Although we evaluate our systems for each concert using the entire reference set, a more realistic scenario should include a significantly larger reference set. For this, we gradually expand our reference set using the MTG-Jamendo dataset (MJD)~\cite{bogdanov2019mtg}, which contains the full audio tracks of 55.7\,k royalty-free songs. We assume that there is no intersection between \datasetname\ and MJD. Due to computation requirements (see Sec.~\ref{sec:run-time}), we only evaluate the Re-MOVE (120,30) setting in this scenario. Table~\ref{tab:large-scale} shows that an increase in reference set size negatively affects the system accuracy. However, the system can still correctly identify 70\% of the annotated segments (before the classifier), and the decrease in performance seems to be saturating after 45\,k references, at least for the considered size regime.

\begin{table}[tb!]
\begin{center}
\label{tab:ablation}
\begin{tabular}{l C{1.cm} C{1.1cm} C{1.3cm} C{1.3cm}}
\hline\hline
VI Config. & TP & FP & DAP (\%) & DLP (\%) \\
\hline
\multicolumn{5}{c}{\textit{Pop/Commercial}} \\
\hline
R - (120, 30) &   \textbf{238}\,/\,\textbf{207} & \textbf{221}\,/\,40   &  \textbf{87.5}\,/\,\textbf{77.2} &    75.5\,/\,\textbf{72.4} \\
Q - (120, 60) &   231\,/\,192 & 231\,/\,\textbf{27}    &  85.7\,/\,72.1 &    \textbf{76.8}\,/\,71.1 \\

\hline
\multicolumn{5}{c}{\textit{Rock/Metal}} \\
\hline
R - (120, 30) &   \textbf{325}\,/\,\textbf{265} & 422\,/\,56   &  83.3\,/\,\textbf{70.7} &    70.1\,/\,66.0 \\
Q - (120, 60) &   321\,/\,263 & \textbf{341}\,/\,\textbf{22}    &  \textbf{83.9}\,/\,70.2 &    \textbf{72.4}\,/\,\textbf{66.9} \\

\hline
\multicolumn{5}{c}{\textit{Indie/Alternative}} \\
\hline
R - (120, 30) &   \textbf{193}\,/\,\textbf{153} & \textbf{118}\,/\,22   &  \textbf{91.8}\,/\,\textbf{73.6} &    73.3\,/\,\textbf{67.3} \\
Q - (120, 60) &   191\,/\,148 & 134\,/\,\textbf{6}    &  89.4\,/\,71.2 &    \textbf{74.0}\,/\,66.0 \\

\hline
\multicolumn{5}{c}{\textit{Hip-hop/Rap}} \\
\hline
R - (120, 30) &   \textbf{76}\,/\,\textbf{60} & \textbf{246}\,/\,47   &  \textbf{43.9}\,/\,\textbf{35.1} &    \textbf{19.7}\,/\,\textbf{17.8} \\
Q - (120, 60) &   63\,/\,47 & 264\,/\,\textbf{16}    &  36.8\,/\,27.5 &    15.6\,/\,13.6 \\

\hline
\multicolumn{5}{c}{\textit{Electronic}} \\
\hline
R - (120, 30) &   \textbf{104}\,/\,\textbf{86} & \textbf{68}\,/\,12   &  \textbf{87.0}\,/\,\textbf{74.8} &    68.5\,/\,64.9 \\
Q - (120, 60) &   99\,/\,\textbf{86} & 69\,/\,\textbf{4}    &  85.2\,/\,\textbf{74.8} &    \textbf{69.6}\,/\,\textbf{66.4} \\

\hline\hline
\end{tabular}
\caption{Results based on genre. R and Q denote Re-MOVE and Qmax, respectively. The left/right values denote the metrics before/after the classifier.}\label{tab:gen-results}
\end{center}
\tabspace
\end{table}

\begin{table}[tb!]
\begin{center}

\begin{tabular}{l C{1.cm} C{1.1cm} C{1.3cm} C{1.3cm}}
\hline\hline
Extra refs. & TP & FP & DAP (\%) & DLP (\%) \\
\hline
None &   936\,/\,771 & 1075\,/\,177   &  80.3\,/\,67.8 &    60.5\,/\,56.9 \\
15k &   860\,/\,678 & 1606\,/\,220   &  73.6\,/\,59.5 &    53.0\,/\,48.8 \\
30k &   836\,/\,661 & 1738\,/\,217   &  71.6\,/\,58.1 &    51.6\,/\,47.4 \\
45k &   812\,/\,643 & 1785\,/\,241    &  69.8\,/\,56.6 &    49.9\,/\,45.9 \\
55.7k &   812\,/\,639 & 1841\,/\,244   &  69.7\,/\,56.2 &    49.6\,/\,45.5 \\

\hline\hline
\end{tabular}
\caption{Results of Re-MOVE (120,30) on the MJD-expanded task.}\label{tab:large-scale}
\end{center}
\vspace{-5mm}
\end{table}

\subsection{Runtime comparison}\label{sec:run-time}

Finally, we share our observations regarding algorithm runtimes. Since computation requirements depend on $W$ and $H$, we here consider $W = 120$ and $H = 30$. Using pre-extracted cremaPCP features as input for each system and executing parallel computations with 32 cores, the Qmax algorithm takes approximately 20\,days to complete the entire distance computations used for the main results in Table~\ref{tab:overall-res}. Contrastingly, both Re-MOVE and 2DFTM take only 11\,min. For the full MJD-expanded task in Table~\ref{tab:large-scale}, the runtime of Re-MOVE only increases to 22\,min (using pre-computed embeddings). Although Re-MOVE and Qmax result in similar performances, the drastic difference in their runtimes suggests that Re-MOVE is the only considered system that both scales up to large-scale retrieval scenarios and achieves a plausible accuracy.

\section{Conclusion and Learnings}
In this work, we have investigated the effectiveness of VI systems for automatic SLI in a wide range of use cases. For this, we have proposed an end-to-end workflow to identify the metadata and timestamps of the songs that are present in full concerts. For the retrieval step, we have compared three VI systems in terms of accuracy and scalability. We have proposed a series of post-processing steps that consolidates and revises the initial retrieved matches to filter out possible false positives for the final results. We have used a new dataset that contains 99.5\,h of concerts, which we publicly share. Our findings suggest that while the audio quality of queries does not have a crucial effect on performance due to the robustness of our input representation against noise, the changes in musical styles/genres can have a drastic impact as our system depends solely on the harmonic information from the audio. For processing the audio queries, using pre-determined window and hop sizes results in imprecise timestamps for the retrieved matches. We have also shown that increasing the size of the reference database negatively impacts the system accuracy. Finally, the reported runtimes for the considered configurations show a remarkable difference between using alignment-based or embedding-based VI systems. Overall, using Re-MOVE for retrieval yields promising results towards automatic SLI in large-scale contexts; however, further improvements for the general workflow are required to address real-world live performance monitoring use cases. In future work, we plan to investigate using an ensemble of VI systems that use various musical characteristics (e.g.,~melody, harmony) for the retrieval phase, and more elaborate false positive filtering schemes.

\clearpage
\bibliographystyle{IEEEbib}
\bibliography{strings,refs}

\end{document}